# Behavioural Sociology of Humans and Machines

Taha Yasseri[a,b]*, Hao Cui[a], Nico Mutzner[c]

[a] Centre for Sociology of Humans and Machines, Trinity College Dublin and Technological University Dublin, Dublin, Ireland
[b] School of Mathematics and Statistics, University College Dublin, Dublin, Ireland
[c] Department of Sociology, University of Zurich, Zurich, Switzerland

All authors contributed equally.
*Corresponding author: taha.yasseri@tcd.ie

**Abstract:**

Artificial intelligence (AI) increasingly participates in communication, decision-making, and collective action, making the social organisation of human–machine interaction central to Hybrid Intelligence (HI). This chapter reviews the behavioural sociology of humans and machines through an HI lens, examining how interaction and feedback shape the combination of human and artificial capabilities. Drawing on sociology, social psychology, human–computer interaction, organisational studies, collective intelligence, and science and technology studies, it traces processes from individual social perception to cooperation in hybrid groups and organisational and institutional arrangements. Particular attention is given to how machine participation alters familiar mechanisms: social categorisation attaches expectations to designed identities, reciprocity may involve unequal social obligations, and algorithmic recommendations acquire authority through organisational practices. The chapter distinguishes observable machine behaviour from human interpretations of intention, understanding, and moral responsibility. It also considers how behavioural effects can extend beyond direct human–machine encounters, while separating established findings from proposed longer-term consequences. It argues that the benefits of HI depend on how roles, incentives, reliance, and accountability are organised, and identifies priorities for empirical research on these evolving sociotechnical arrangements.

# 1. Introduction

Artificial intelligence (AI) is no longer confined to laboratories or specialised technical applications. Intelligent systems increasingly participate in everyday social, economic, and political life. They recommend what we read and watch, assist with writing and decision-making, moderate online content, allocate resources, and interact with people in workplaces, educational settings, and public institutions. In many of these settings, AI systems shape how individuals and groups behave, interact, and make decisions.

The growing integration of AI into society has introduced questions that extend beyond computer science. Societal consequences of AI use are not determined solely by algorithmic properties. Technologies enter existing social environments populated by individuals, groups, organisations, norms, and institutions. Once deployed, they are adopted, resisted, adapted, and repurposed, often in ways that differ substantially from the intentions of their designers. Digital technologies provide many examples of this process. Social media platforms developed for communication and connection have become important infrastructures for political mobilisation, collective action, and other forms of political participation (Tufekci & Wilson, 2012; Earl & Kenski, 2022). Digital platforms have also created new opportunities for citizen participation and engagement in democratic decision-making (Fischli & Muldoon, 2024). Similarly, recommendation systems designed to support information discovery have transformed cultural consumption, public discourse, and political communication. Understanding the societal implications of AI requires attention to technological capabilities and to the social contexts in which these technologies are used.

Across several disciplines, scholars increasingly recognise that the participation of artificial agents raises new questions for sociological inquiry. Tsvetkova et al. (2024) argue for a sociology of humans and machines that treats human–machine systems as a distinct object of study. In this chapter, behavioural sociology of humans and machines refers to the study of how interactions among humans and artificial agents generate behavioural, organisational, and societal outcomes. Artificial agents include conversational agents, recommender systems, decision-support systems, and autonomous systems whose outputs or actions elicit human responses. Hybrid Intelligence (HI) seeks to combine human and artificial capabilities to achieve complex goals, drawing on complementarity and learning through interaction (Dellermann et al., 2019). We approach HI by examining the social processes through which these capabilities become interdependent. Complementarity is an empirical achievement: it depends on how people interpret machine behaviour, allocate tasks, and organise reliance and responsibility. The same interdependence can also produce exploitation, correlated errors, or weakened cooperation. Our distinctive focus is on how machine participation changes the operation of familiar social mechanisms, including categorisation, reciprocity, and authority, and how these changes affect collective outcomes. This approach connects the study of human–machine interaction to the social conditions under which HI succeeds or fails.

A complementary perspective has emerged from studying machine behaviour. Rahwan et al. (2019) argue that intelligent systems should be investigated as behavioural entities whose actions can be observed, measured, and explained. This perspective becomes increasingly important as machines interact with humans and with other machines. Recommendation algorithms compete for attention, automated trading systems respond to one another in financial markets, and autonomous software agents may even develop persistent conflicts, as illustrated by long-running interactions among Wikipedia bots (Tsvetkova et al., 2017a). Therefore, understanding machine behaviour has become an essential complement to understanding human behaviour and the societies in which they coexist.

Much of the existing literature has focused on dyadic interactions between a single human and a single intelligent system. This work has generated important insights into anthropomorphism, trust, and human-computer interaction. However, many contemporary human-machine systems operate at much larger scales. People interact with multiple AI systems simultaneously; AI systems mediate interactions among many people; and collective outcomes emerge through networks of both human and artificial actors. Sociological analysis operates at this larger scale and can explain how interactions with AI affect social relationships, group dynamics, organisational processes, and institutional structures. These transformations may occur indirectly through changes in communication patterns, information flows, social norms, and collective decision-making.

This chapter provides a narrative review of the sociobehavioural foundations of human–machine systems. Drawing on research from sociology, behavioural science, human-computer interaction, organisational studies, collective intelligence, and science and technology studies, it examines how anthropomorphism, social categorisation, bias, trust, fairness, cooperation, and other social processes shape human–machine interaction and its broader consequences. The chapter brings these different strands of research together through the lens of Hybrid Intelligence, with particular attention to the interplay between human and machine behaviour and the feedback that develops through their interaction. This approach allows us to examine how human–machine interdependence develops across different levels of social organisation. In doing so, the chapter considers both what a behavioural sociological perspective can contribute to the study of Hybrid Intelligence and what Hybrid Intelligence means for established sociological concepts and explanations.

# 2. Conceptual and Methodological Foundations

The behavioural sociology of humans and machines draws on several established disciplines. Although its central questions are sociological, explicitly sociological research in this area remains relatively limited, and much of its theoretical and methodological foundation comes from neighbouring fields. These perspectives differ in their assumptions, units of analysis, and methodological approaches, offering different ways of studying human–machine interaction and its social consequences. This section reviews the theoretical perspectives and methodological approaches that underpin this research.

## 2.1 Theoretical Perspectives

No single discipline fully explains behaviour in human-machine systems. Human-computer interaction explains how people engage with computational systems; social psychology identifies the cognitive and behavioural mechanisms underlying these interactions; sociology examines how they scale into norms, institutions, and social structures; organisational studies investigate coordination and decision-making within formal organisations; collective intelligence explores how groups solve problems collectively; and science and technology studies emphasises the co-evolution of technologies and societies. Together, these perspectives offer complementary insights for understanding how behaviour emerges within systems composed of both human and artificial actors.

### Human–Computer Interaction

Human-computer interaction (HCI) provides one of the earliest and most influential traditions for studying interactions between people and computational systems. Initially focused on usability, interface design, and performance (Card et al., 1983), the field later showed that people often respond to computers as if they were social actors. The Computers Are Social Actors (CASA) paradigm showed that users apply norms of politeness, reciprocity, and personality judgements even while recognising that they are interacting with machines (Nass et al., 1994; Nass & Moon, 2000). HCI therefore established that interaction with technology depends not only on technical performance but also on the social expectations users bring to these encounters. This work laid the foundations for subsequent research on human-AI interaction.

## Social Psychology

If HCI established that people respond socially to intelligent systems, social psychology explains why. People interpret limited social cues to form impressions of others, attribute intentions and capabilities, and place interaction partners into socially meaningful categories (Asch, 1946; Heider, 1958; Fiske & Taylor, 2013). These interpretations shape expectations about characteristics such as competence, warmth, and trustworthiness, which can in turn influence subsequent behaviour. As AI systems become increasingly conversational and human-like, similar processes can shape how people perceive and respond to artificial agents. Social psychology therefore provides a basis for understanding how perceptions of AI develop and how they influence trust, bias, reciprocity, and cooperation in human–machine interaction.

## Sociology

Sociology examines how interactions are embedded in social relations and contribute to collective, organisational, and institutional outcomes. Sociotechnical systems research has long treated work outcomes as depending on the organisation of both social and technical processes, directing attention to how technology is integrated with roles, autonomy, and task design (Trist & Bamforth, 1951; Parker & Grote, 2022). Sociomaterial perspectives likewise challenge separating technologies from the practices through which they are developed and used (Scott & Orlikowski, 2025). These traditions establish a foundation for analysing human–machine interdependence. Recent work on the sociology of humans and machines builds on this foundation by examining artificial agents that participate directly in communication, coordination, and collective behaviour (Tsvetkova et al., 2017b; Tsvetkova et al., 2024; Yasseri, 2025). A behavioural perspective asks how responses to machine behaviour alter cooperation, distributions of influence, and norm formation, and how organisational arrangements shape those responses. For HI, it connects the combination of capabilities to the social conditions of their use.

## Organisational Studies

Organisations provide important settings for studying how human–machine interaction becomes embedded in work practices and formal structures. Introducing AI can alter how tasks are allocated, how work is coordinated, and how responsibilities are distributed across organisational roles (Yasseri et al., 2026). Work design research shows that these effects depend on how technologies are integrated into the social and organisational conditions of work, including autonomy, skill use, feedback, and interaction patterns (Parker & Grote, 2022). Research on human–AI teams also highlights communication, coordination, and shared mental models as processes through which humans and artificial agents develop expectations about one another's capabilities and responsibilities (Andrews et al., 2023).

## Collective Intelligence

Collective intelligence examines how interactions among individual members give rise to collective problem solving and performance. Research in this stream has identified diversity, communication, coordination, and the distribution of expertise as important conditions shaping collective outcomes (Woolley et al., 2010; Malone & Bernstein, 2015). The increasing participation of AI extends these questions to groups in which both human and artificial agents may contribute information and make decisions (Cui & Yasseri, 2024). From a Hybrid Intelligence perspective, this directs attention to how human and machine contributions are combined through interaction and how their interdependence shapes outcomes at the group level.

## Science and Technology Studies

Science and technology studies (STS) remind us that technologies do not possess fixed social consequences determined solely by their technical design. Instead, technologies and societies continually shape one another through processes of development, adoption, adaptation, and governance (Bijker et al., 1987; Latour, 2005). This perspective directs attention beyond algorithms to the broader sociotechnical systems within which human-machine interactions occur. Questions of trust, fairness, cooperation, and accountability are therefore understood as properties of evolving social systems rather than of technologies alone. STS thus provides an important foundation for understanding how behavioural, organisational, and institutional processes co-evolve with increasingly intelligent technologies.

These theoretical perspectives approach human–machine systems from different units and levels of analysis. Some focus on how people perceive and respond to machines, others on coordination and collective performance, and still others on the organisational and institutional settings in which these interactions become embedded. The Hybrid Intelligence perspective connects these levels by focusing on the interactions through which human and machine behaviour becomes interdependent. For behavioural sociology, this provides a basis for tracing how interactions between humans and machines become connected to broader patterns of collective behaviour, organisational practice, and institutional change.

These perspectives suggest three ways in which machine participation may change familiar social mechanisms. First, categorisation can attach gendered or status-based expectations to identities deliberately assigned through names, voices, and organisational roles. Second, reciprocity can involve unequal obligations: people may expect an artificial partner to cooperate while feeling less obliged to return its cooperation. Third, authority can attach to machine-generated outputs through delegation and organisational rules, even where the system has no corresponding human understanding of its role. In each case, an observable pattern familiar from human interaction may involve different processes or a different distribution of obligations and influence. These possibilities provide questions for empirical investigation rather than grounds for assuming that machines possess human intentions or moral commitments. The following sections examine the evidence for these changes and their consequences for hybrid groups and organisations.

## 2.2 Methodological Approaches

The behavioural sociology of humans and machines draws on methods suited to different levels of analysis. Experiments can identify causal effects of system characteristics and interaction conditions, surveys capture perceptions and attitudes, and digital trace data and field studies provide evidence of behaviour in natural settings. Computational simulations offer another way to examine collective dynamics that are difficult to observe directly.

Similar methods can serve different purposes across disciplinary perspectives. HCI experiments often manipulate interface features or system behaviour to examine how people interact with and respond to a technology. Social psychological experiments use controlled manipulations to examine mechanisms such as attribution, categorisation, trust, and social influence. Organisational and collective intelligence research extends experimental designs to questions of coordination, task allocation, team composition, and group performance, while sociological and STS research more often draws on field studies and case studies to examine interactions within broader social and institutional settings. Methodological differences across these perspectives concern the processes and levels of analysis that research designs aim to capture.

Studying human–machine systems also requires attention to interaction over time. Human behaviour can affect subsequent machine responses, which may in turn alter human expectations, decisions, and coordination patterns. From a Hybrid Intelligence perspective, this makes repeated interaction, feedback, and adaptation important objects of empirical study. Research designs may consequently need to capture human behaviour, machine behaviour, and changes in their interaction across multiple rounds or longer periods.

Research on human–AI teaming highlights the importance of specifying team composition, task structure, interaction processes, and both individual and collective outcomes (Flathmann et al., 2025; Schmutz et al., 2024). Example studies show how these choices allow researchers to distinguish responses to an agent's behaviour from responses to its perceived identity, and to examine changes across repeated encounters.

Ishowo-Oloko et al. (2019), for example, studied a repeated prisoner's dilemma in which participants interacted with a human or a bot and received true or false information about their partner's identity. Crossing actual and reported identity allowed the researchers to distinguish machine performance from the effects of being perceived as artificial. Bots were more effective at eliciting cooperation, but disclosure of their identity removed this advantage. The study shows how social interpretation can obstruct the translation of technical capability into cooperative outcomes.

Shirado et al. (2023) examined pairs of participants remotely driving robotic vehicles in a repeated coordination task. They varied braking and steering assistance and the availability of communication. Tracking concessions over successive rounds, and behaviour after assistance was deactivated, revealed effects on reciprocity that would be missed by measuring immediate task performance alone.

These designs address different explanatory questions. Manipulating perceived identity helps identify categorisation effects; observing sequences of interaction helps identify adaptation and behavioural carryover. Studies of organisations additionally need to examine how rules, incentives, and opportunities to contest decisions shape these processes. Table 1 summarises the principal perspectives, units of analysis, research questions, and methods. The next section examines how social perception influences human–machine interaction.

**Table 1.** Principal theoretical perspectives informing the behavioural study of human–machine systems, with their primary units of analysis, research questions, and methodological approaches.

| Discipline | Primary Unit of Analysis | Main Questions | Common Methods |
|---|---|---|---|
| Human–Computer Interaction | Individual | How do humans perceive and interact with AI systems? | Experiments, Usability Studies |
| Social Psychology | Individual | How do cognitive and social processes shape human–AI interaction? | Experiments, Surveys |
| Sociology | Groups & Society | How do AI systems reshape norms, roles, and social relations? | Surveys, Field Studies, Digital Trace Data |
| Organisational Studies | Organisations | How do humans and AI coordinate work within organisations? | Experiments, Field Studies |
| Collective Intelligence | Groups | Under what conditions do groups of humans and AI become collectively intelligent? | Experiments, Digital Trace Data, Simulations |
| Science and Technology Studies | Sociotechnical Systems | How do AI technologies and societies co-evolve? | Field Studies, Case Studies |

# 3. Social Perception and Bias in Human–Machine Systems

Human–machine interaction depends on what machines do and on how humans respond to their behaviour. Anthropomorphism and social categorisation provide familiar cognitive and social frames through which people make sense of artificial agents, attribute characteristics to them, and form expectations about their behaviour. These processes can also introduce or reinforce biases in how people evaluate machines and their outputs. Together, anthropomorphism, social categorisation, and bias link interpretations of machine behaviour to subsequent patterns of trust, cooperation, authority, and collective behaviour. This section examines these processes and their implications for human–machine systems.

## 3.1 Anthropomorphism

Anthropomorphism refers to the attribution of human characteristics, including intentions, emotions, agency, personality, understanding, and moral capacity, to non-human entities (Epley et al., 2007; Gray et al., 2007). Within human-machine systems, anthropomorphism emerges through the interaction between system design and human cognition. Designers incorporate cues such as conversational language, synthetic voices, avatars, names, role descriptions, first-person expressions, and adaptive behaviour, while users interpret these cues using cognitive processes that evolved for social interaction with other people (Li & Suh, 2022; Seeger et al., 2017). As a result, people often perceive AI systems in social rather than purely technical terms.

This tendency has been recognised for several decades within human-computer interaction. Research consistently demonstrates that people readily apply norms of politeness, reciprocity, personality judgements, and other social expectations to computational systems, even when they know they are interacting with machines (Nass et al., 1994; Nass & Moon, 2000). More recent studies extend these findings across chatbots, virtual assistants, recommender systems, and service robots, showing that anthropomorphic design influences perceived warmth, competence, trust, acceptance, and willingness to cooperate (Blut et al., 2021; Li & Suh, 2022). Collectively, this literature suggests that anthropomorphism is not an exceptional response but a common feature of human interaction with increasingly sophisticated technologies.

The emergence of large language models (LLMs) has made anthropomorphism particularly salient. Fluent natural-language communication can make these systems appear intentional, knowledgeable, or socially aware, but conversational fluency alone does not establish human-like understanding or mental states (Abercrombie et al., 2023; Shanahan, 2024). Interface design can further amplify these perceptions. For example, Cohn et al. (2024) found that combining speech with text increased perceived anthropomorphism, while first-person language influenced perceptions of accuracy and risk in some contexts. These findings illustrate that anthropomorphic responses depend not only on underlying models but also on how AI systems are presented to users.

Anthropomorphism also plays an important role in embodied and autonomous systems. Attributing a "mind" to an autonomous vehicle has been shown to increase trust in its operation (Waytz et al., 2014). Similar effects have been observed for social robots and service agents that appear responsive, intentional, or socially aware (Blut et al., 2021; Duffy, 2003). Although these perceptions may facilitate cooperation and user acceptance, they can also encourage people to overestimate machine understanding, emotional capacity, autonomy, or moral responsibility (Lee & See, 2004). Anthropomorphism therefore offers an opportunity to improve human-machine interaction but can also lead to inappropriate reliance.

From a behavioural sociological perspective, anthropomorphism is important because it shapes how people define social situations involving intelligent systems (Goffman, 1974; Suchman, 1987). Human-like cues encourage users to draw on familiar social expectations regarding agency, competence, responsibility, trustworthiness, and care when interacting with AI. These interpretations influence not only individual behaviour but also patterns of cooperation, authority, and responsibility within broader human-machine systems (Dellermann et al., 2021). Anthropomorphism thus provides the behavioural foundation for many of the processes discussed throughout this chapter. Building on this perspective, the next section examines how social categorisation organises expectations around gender, race, and status, and how these expectations differ from trait impressions such as warmth and competence.

## 3.2 Social Categorisation

Social categorisation refers to the tendency to organise social perception through meaningful social groups and roles (Allport, 1954; Tajfel et al., 1971). Rather than seeing others as entirely unique individuals, people routinely classify them by categories such as gender, race, class, occupation, status, and age. These categories enable rapid social judgement by letting people infer likely behaviours, intentions, and social roles (Macrae & Bodenhausen, 2000). At the same time, they carry culturally shared stereotypes, norms, and status hierarchies that shape expectations before direct interaction occurs (Fiske, 1998; Ridgeway, 2014). Social categorisation therefore provides a cognitive shortcut that facilitates social interaction while also creating the conditions for stereotypes and biases to emerge.

Gender illustrates how these expectations can transfer to artificial agents (Eyssel & Hegel, 2012). Role congruity theory links evaluations of leaders to the fit between gender stereotypes and role expectations (Eagly & Karau, 2002). Cui and Yasseri (2026) examined this process in teams working under managers described as human or AI and as male, female, or gender-unspecified. Evaluations depended on the award outcome, with female AI managers facing particularly negative judgements when participants did not receive an award. The relevant mechanism is the interpretation of an assigned identity in a consequential interaction. A system's presentation can therefore reproduce gendered leadership expectations without gender being an intrinsic property of its computational operations. Assigned gender has also been found to affect cooperative decisions in human–AI interaction (Bazazi et al., 2025), connecting social categorisation to behaviour as well as evaluation.

Status expectations also shape how people respond to artificial agents. In human groups, status characteristics generate expectations of competence and influence (Berger et al., 1972; Ridgeway, 2014). In human–machine settings, role labels such as assistant, adviser, or manager may position the same technology differently within an interaction. Earlier experiments showed that interdependence with a computer could affect perceived similarity and openness to its influence (Nass et al., 1996). For HI, this directs attention to how role assignments and institutional settings organise whose contributions are heard and followed. Perceived status should also be distinguished from the formal authority conferred by organisational procedures, discussed in Section 5.2.

Social identity categories should be distinguished from trait impressions such as warmth and competence. These impressions help people anticipate whether an interaction partner appears capable and aligned with their interests (Fiske et al., 2018). McKee et al. (2024) studied human interaction with reinforcement-learning agents in a social dilemma. Perceived warmth and competence predicted preferences for agents beyond objective performance measures, including choices about whether to continue interacting. This suggests that the selection of artificial partners depends on social evaluation as well as task capability, with consequences for how hybrid groups form and sustain collaboration.

Race and intersectionality remain important for understanding these processes. Racial classification organises perceptions of belonging and hierarchy, while its effects intersect with gender and other social positions (Omi & Winant, 2014; Bonilla-Silva, 1997; Collins, 2022; Crenshaw, 1989, 1991). In human–machine interaction, research needs to establish which design and communication cues activate such classifications and how their effects vary across users and contexts. This question is distinct from whether a model's outputs discriminate against human demographic groups. The former concerns how people categorise an artificial agent; the latter concerns unequal treatment of people through a sociotechnical system, examined in Section 3.4.

Social categorisation matters for HI because it influences expectations about who should contribute, who should be trusted, and whose decisions should carry weight. These expectations can affect task allocation and cooperation before differences in actual capability have been established. The analytical task is to distinguish the effects of assigned identity, observed behaviour, and organisational position, and to examine how their interaction shapes collective outcomes.

## 3.3 Bias in Human Judgements of AI

Bias in human judgements of AI refers to the systematic ways in which users evaluate machine outputs, capabilities, and social roles differently across contexts. These judgements extend beyond assessments of technical performance to include perceptions of trustworthiness, competence, authority, objectivity, legitimacy, and responsibility. They are shaped not only by the AI system's characteristics, but also by how it is presented, the task it performs, the degree of uncertainty involved, and the social expectations users bring to the interaction (Glikson & Woolley, 2020; Lee & See, 2004). Building on the processes of anthropomorphism and social categorisation discussed above, this section examines how these perceptions translate into systematic biases in human judgement.

One of the most extensively studied forms of bias concerns the calibration of trust and reliance. Users may place excessive confidence in automated recommendations, a phenomenon commonly called *automation bias* (Goddard et al., 2012). Conversely, they may discount algorithmic advice after observing a small number of errors, leading to *algorithm aversion* (Dietvorst et al., 2015). Under other conditions, users may prefer algorithmic over human judgement, particularly in analytical and predictive tasks where machines are perceived as more objective or consistent, a pattern described as *algorithm appreciation* (Logg et al., 2019). Together, these findings demonstrate that reliance on AI depends not only on objective performance but also on users' beliefs about the appropriate role of machine judgement.

These evaluations also vary across social and moral domains. People generally express greater acceptance of AI in tasks framed as calculation, optimisation, or prediction than in decisions involving healthcare, criminal justice, or moral evaluation (Castelo et al., 2019;

Longoni et al., 2019). Research on machine moral decision-making similarly suggests that users are reluctant to accept AI as a legitimate moral decision-maker because intelligent systems are perceived as lacking essential human qualities such as empathy, intentionality, and moral understanding (Bigman & Gray, 2018). These findings indicate that perceptions of legitimacy depend not only on technical competence but also on beliefs about which forms of judgement require uniquely human capacities.

The categorisation processes discussed in Section 3.2 can become evaluative biases when assigned identities influence judgements independently of relevant behaviour or capability. For HI, their significance lies in the consequences for participation and influence: people may discount useful contributions, defer to less capable systems, or evaluate equivalent decisions differently. The effects of such biases extend to the humans working within or affected by these arrangements. A gendered evaluation of an artificial manager, for example, may reproduce broader leadership stereotypes and shape acceptance of organisational decisions.

From a behavioural sociological perspective, these biases extend well beyond individual judgement. They influence whether people accept or challenge AI recommendations, how authority is distributed within hybrid groups, and how people attribute responsibility for decisions and outcomes. Bias therefore links social perception and collective behaviour, shaping how intelligent systems integrate into organisations, institutions, and everyday social life. The following section examines how these judgements interact with data, models, and organisational practices to embed bias within wider human–machine systems.

## 3.4 Bias Embedded in Human-Machine Systems

Bias becomes embedded in human-machine systems when patterns originating in data, social assumptions, and human behaviour become stabilised across sociotechnical systems. Rather than residing solely within algorithms, bias emerges through interactions among models, users, organisations, and institutions. It may enter during data collection, annotation, model development, interface design, or deployment (Suresh & Guttag, 2021), but it can also arise after deployment as people interpret, accept, reject, or act upon AI system outputs. Consequently, bias should be understood as a property of the broader sociotechnical system rather than of individual algorithms alone (Selbst et al., 2019).

At the technical level, AI systems frequently learn from historical data that reflect existing inequalities in domains such as recruitment, policing, lending, healthcare, and education (Fabris et al., 2025; Ferrara, 2024). Models trained on these data may reproduce or amplify historical disparities by converting unequal social patterns into predictive signals. Even when protected characteristics are excluded, variables such as postcode, educational background, employment history, language use, or consumption patterns may act as proxies, indirectly generating discriminatory outcomes (Barocas & Selbst, 2016). Friedman and Nissenbaum (1996) distinguish between pre-existing, technical, and emergent bias, highlighting that bias may originate in social conditions, system design, or patterns of use. Subsequent empirical work has demonstrated how dataset composition and model performance can produce substantial disparities across demographic groups, particularly at the intersection of gender and race (Buolamwini & Gebru, 2018).

Bias becomes further embedded through organisational practice and collective behaviour. Algorithmic recommendations may be interpreted as objective evidence, selectively accepted, or incorporated into routine decision-making, allowing local judgements to acquire institutional authority. Research on algorithm-in-the-loop decision-making demonstrates that fairness depends not only on model performance but also on how people interpret, question, and integrate algorithmic advice into organisational practice (Green & Chen, 2019). These interactions often generate feedback loops. Decisions influenced by AI become new training data, which subsequently shape future algorithmic recommendations, reinforcing existing behavioural patterns and organisational routines. Over time, these recursive processes may normalise discriminatory outcomes by making them appear efficient, objective, or evidence-based.

From a behavioural sociological perspective, bias is therefore best understood as an emergent property of human-machine systems rather than a defect of individual models. Addressing it requires interventions across the entire sociotechnical system, including data governance, model evaluation, interface design, user training, institutional oversight, and the continuous monitoring of social outcomes (Raji et al., 2020). Responsibility for mitigating bias is similarly distributed across designers, organisations, regulators, and users. This systems perspective provides an important transition to the next section, which examines how trust, cooperation, and coordination emerge once humans and intelligent systems interact repeatedly within groups and organisations.

# 4. Cooperation and Collective Behaviour in Hybrid Groups

Cooperation is a fundamental organising principle of social life and one of the domains in which artificial agents most visibly reshape collective behaviour. Much of what is currently known about human-AI cooperation comes from dyadic interactions between a single human and a single machine, largely studied within the human-computer interaction tradition. This work has generated important insights into trust, reliance, and decision-making, providing the first evidence of how humans respond to increasingly capable AI systems. However, contemporary human–machine systems increasingly involve interactions within hybrid groups, where artificial agents can affect individual behaviour as well as relationships and coordination among people. Understanding cooperation in these settings requires moving beyond the dyad to consider group and collective processes.

Recent reviews of AI-enhanced collective intelligence argue that human and AI capabilities can complement one another, allowing human–AI systems to achieve forms of collective intelligence that neither humans nor AI could achieve in isolation (Cui & Yasseri, 2024), aligning with ideas of Hybrid Intelligence. In the context of cooperation, a Hybrid Intelligence perspective highlights how human and machine behaviour becomes interdependent through interaction and feedback, and how this interdependence affects collective outcomes. Recent work in computational sociology similarly examines how interactions among human and artificial agents give rise to patterns of cooperation and conflict in hybrid societies (Yasseri, 2025).

## 4.1 Human–AI Cooperation

Cooperation is fundamentally a problem of social order. Human societies sustain cooperation through shared norms of reciprocity, trust, and fairness that enable mutually beneficial interaction. Introducing an artificial agent can alter this equilibrium because these norms often apply differently to machines than to humans. The growing literature suggests that AI neither consistently erodes nor consistently strengthens cooperation (March, 2021; Chugunova & Sele, 2022). Instead, its effects depend on a combination of behavioural, organisational, and institutional mechanisms that shape how people perceive and incorporate artificial agents into social interaction.

We use cooperation in a behavioural sense to describe actions that support joint or collective outcomes, including contributions that benefit others and coordination towards shared goals. This usage does not imply equivalent motivations or forms of agency among human and artificial participants. In particular, a machine's contingent response to another actor's behaviour should be distinguished from the actor's interpretation of that response as an expression of social obligation.

A recurrent finding is the machine penalty: people often cooperate less with a partner identified as artificial than with a human partner (Bonnefon et al., 2024; Zhou et al., 2026). Research also reports greater willingness to exploit machine partners and reduced feelings of guilt about doing so (De Melo et al., 2016; Karpus et al., 2021). In Ishowo-Oloko et al. (2019), disclosure of a bot's identity undermined its ability to elicit cooperation despite its cooperative behaviour. These findings suggest that responding to a partner's actions and feeling an obligation to reciprocate are separable components of cooperation.

This distinction changes how reciprocity should be analysed in hybrid systems. An artificial agent can implement a cooperative strategy while human participants assign it weaker claims to reciprocal treatment. Cooperative behaviour may consequently coexist with unequal social obligations. The resulting outcome depends on the agent's strategy, users' expectations, and the incentives governing interaction. A related issue arises when people delegate actions to machines: delegation can make dishonest behaviour easier when an artificial agent carries out instructions that a human proxy might reject (Köbis et al., 2021, 2025). For HI, understanding these asymmetries is necessary for designing arrangements that sustain cooperation over repeated encounters.

AI can also support cooperation under particular conditions. Perceived warmth and competence are associated with preferences for artificial partners and choices about whether to continue interacting, beyond objective performance measures (McKee et al., 2024). These findings complement identity effects: evaluations of an agent's behaviour and beliefs about its artificial status can both shape interaction. Computational evolutionary modelling identifies additional possibilities. Hintze and Adami (2026) found that artificial agents mimicking player behaviour lowered the threshold for cooperation in a modelled public goods game, identifying a mechanism for testing in human–AI groups. Experiments also show that rewards and punishments can reduce biases against cooperation partners assumed to be machines (Makovi et al., 2025), complementing evidence on the role of sanctions in sustaining human cooperation (Fehr & Gächter, 2002). Likewise, delegating decisions to artificial agents can encourage prosocial behaviour under some conditions, even though delegation to AI may promote dishonest behaviour in other contexts (Fernández Domingos et al., 2022; Köbis et al., 2025).

These studies suggest that cooperation depends on the interplay between an interaction partner's identity and behaviour and the social context in which the interaction takes place. Whether AI strengthens or weakens cooperation also depends on the interaction's strategic structure, the artificial agent's position within the group, and the institutional arrangements that govern collective behaviour (Guo et al., 2023). Artificial agents embedded within groups can promote intergroup cooperation by reducing ingroup favouritism and encouraging prosocial behaviour towards outgroup members (Igwe & Durrheim, 2024). Performance gains from AI assistance also vary across tasks. AI-assisted workers can outperform unaided workers on tasks within the system's capabilities while performing worse on tasks beyond them, reflecting the "jagged technological frontier" identified by Dell'Acqua et al. (2026). Effective collaboration requires reliance that is sensitive to the AI's performance on the current task (Lee & See, 2004). Excessive dependence through automation bias and unwarranted rejection through algorithm aversion can both impair decisions; appropriate reliance therefore involves accepting useful advice while retaining the ability to reject incorrect recommendations (Romeo & Conti, 2026; Dietvorst et al., 2015; Burton et al., 2020; Schemmer et al., 2023).

Research on human–AI cooperation increasingly examines the conditions under which different cooperation characteristics and outcomes emerge. From a behavioural sociological perspective, distinct cooperation characteristics develop through the interaction between intelligent systems and the social settings in which they are embedded. This makes cooperation a property of the wider human-machine system, shaped by expectations, norms, incentives, and organisational structures.

## 4.2 Collective Intelligence and Emergent Collective Behaviour

As human–machine systems expand beyond dyadic interactions, a central question is whether hybrid groups can achieve collective intelligence that exceeds the capabilities of humans or AI alone. Research on human collective intelligence highlights how diversity, communication, and knowledge aggregation can help groups compensate for individual limitations (Woolley et al., 2010). Evidence on human–AI performance, however, varies across tasks and comparison baselines. Vaccaro et al. (2024) found that human–AI combinations, on average, outperformed humans alone but performed worse than the better of humans or AI alone. Gains were greater in content-creation tasks, while decision-making tasks showed performance losses. These findings distinguish human augmentation from synergy; their implications for multi-person hybrid groups depend on the interaction structures and tasks being studied.

Many of the conditions identified in the collective intelligence literature remain equally important in hybrid groups. Diversity, independence of judgement, and appropriate mechanisms for aggregating contributions continue to influence collective performance (Hong & Page, 2004). These conditions may be harder to maintain when some group members are AI systems. Recent evidence suggests that using large language models can homogenise linguistic expression and reduce variation across individual contributions (Sourati et al., 2026). In hybrid groups, this raises the possibility that multiple AI-assisted or AI-generated contributions may provide less diversity than their number would suggest. Designing hybrid groups requires attention to the diversity and complementarity of human and machine contributions, since adding AI does not necessarily increase collective diversity.

Beyond collective intelligence lies the broader question of emergent collective behaviour: patterns arising from interactions among many agents (Rahwan et al., 2019; Tsvetkova et al., 2024). Different research designs provide evidence at different levels. Agent-based modelling suggests that political bots can amplify opinion polarisation in social networks (Lu & Lee, 2025), while experiments with human–robot teams show that socially expressive robots can improve conversational dynamics among people (Traeger et al., 2020). Research on human populations demonstrates that committed minorities can trigger changes in social conventions (Centola et al., 2018). Ashery et al. (2025) found related convention-formation and minority-influence dynamics in populations composed entirely of LLM agents. A recent preprint likewise reports that conformity can produce collective misalignment among individually aligned AI agents (De Marzo et al., 2026). AI-only studies identify mechanisms and hypotheses relevant to HI, but establishing how these dynamics operate in mixed human–AI populations requires direct investigation.

Artificial agents can also change the norms governing relationships among humans. In Shirado et al. (2023), steering assistance inhibited reciprocal concessions between human participants, and the negative effects persisted after assistance was deactivated. This illustrates how automating part of an interaction can change the behavioural opportunities through which people establish reciprocity. Other research finds persistent perceptual effects following social norm learning from non-human agents (Duderstadt et al., 2022), while trust within hybrid collectives depends on perceived consensus about cooperative norms (Makovi et al., 2023).

Gu et al. (2026) propose a broader mechanism of norm leakage: interaction habits acquired with conversational AI may carry over into human–human encounters. Their perspective reviews emerging evidence and suggests that sycophantic systems may amplify this process; it does not establish the durability or societal extent of the proposed effects. Analytically, behavioural carryover should be distinguished from institutionalisation. Persistence after an encounter is evidence of an effect beyond the immediate dyad, whereas the emergence of a stable social norm additionally requires shared expectations and reinforcement across interactions. Establishing that progression calls for longitudinal research connecting repeated human–AI encounters to subsequent human relationships and organisational routines.

These studies show how artificial agents increasingly participate in the behavioural processes through which collective intelligence, social norms, and other group-level outcomes develop. Their effects can also extend beyond direct human–machine interaction by changing relationships and coordination patterns among people. Understanding these emergent dynamics requires extending the analysis from dyadic human–AI interaction to groups, organisations, and societies as evolving human-machine systems. This shift from isolated human-AI interactions towards hybrid societies reflects a broader transition in computational social science, where increasingly autonomous machines become participants in collective social processes rather than simply technological artefacts (Yasseri, 2025; Tsvetkova et al., 2024).

# 5. Behavioural and Institutional Dynamics of Human–Machine Systems

The preceding sections examined how humans perceive and interact with artificial agents and how these interactions shape cooperation and collective behaviour. As human–machine interaction becomes embedded in organisations and institutions, repeated patterns of reliance, coordination, and delegation can develop into more stable roles, routines, and expectations. These arrangements raise questions of trust, fairness, authority, and legitimacy, as well as how responsibilities and influence are distributed between human and machine participants. Over time, they may become incorporated into organisational practices, shared norms, and institutional structures. This section examines these processes and how organisational and institutional arrangements, in turn, shape subsequent human–machine interaction.

## 5.1 Trust, Fairness, and Legitimacy

Trust helps organise cooperation in human–machine systems, but should be distinguished from observed reliance on an AI system. Users may regard a system as trustworthy while declining its advice on a particular task, or follow its outputs because organisational procedures constrain their alternatives. Research on trust in AI examines perceptions of competence, reliability, predictability, and alignment with human goals, as well as the influence of social cues and context (Lee & See, 2004; Glikson & Woolley, 2020). These perceptions do not establish that a machine has the intentions or moral commitments users attribute to it. From an HI perspective, research needs to connect reported trust to behaviour, including the acceptance, checking, and rejection of outputs. Trust can develop through repeated interaction and change in response to system behaviour and context (Glikson & Woolley, 2020; Liebherr et al., 2026).

A central insight emerging from this literature is that effective human–AI collaboration depends on appropriate trust calibration rather than simply maximising trust. Excessive reliance can lead users to accept AI outputs without sufficient scrutiny, while unwarranted scepticism can lead them to discount useful algorithmic advice. Research on explanations examines their effects on users' responses to correct and incorrect advice; these effects should be evaluated through observed reliance rather than assumed to represent trust repair following errors (Schemmer et al., 2022, 2023). These findings locate reliance in the interaction between system behaviour, human expectations, and organisational context.

Closely related to trust are perceptions of fairness. In human–AI systems, fairness concerns the treatment of people and groups and the procedures through which decisions affecting them are made. Transparency, explainability, contestability, and accountability can support these procedures, but none establishes that outcomes are fair. Users' perceptions of fairness can influence whether they accept AI recommendations, while the effects of transparency and explanations on trust and reliance depend on context (Glikson & Woolley, 2020). Fairness also extends beyond algorithmic accuracy. It requires examining how recommendations are interpreted and used, whose interests are served, and which human and organisational actors are accountable for decisions (Green & Chen, 2019; Raji et al., 2020; Selbst et al., 2019). Fairness should therefore be understood as a property of the broader sociotechnical process, including the model, interface, users, organisational procedures, and accountability structures.

Trust and fairness together contribute to the broader question of legitimacy. Legitimacy concerns whether people regard AI as an appropriate actor to perform particular social roles and exercise influence over others. Acceptance of AI in these roles depends on users' expectations about responsibility, reciprocity, and social obligation. Humans often expect AI to behave fairly and cooperatively while feeling weaker obligations to reciprocate, placing AI in an intermediate social position between tool and social partner (Harrell & Traeger, 2025; Karpus et al., 2021). These asymmetries suggest that legitimacy is negotiated socially rather than determined solely by technical capability.

From a behavioural sociological perspective, trust, fairness, and legitimacy can be understood as relational processes that stabilise and organise human–machine relationships. They develop through repeated encounters between system behaviour and human expectations, while shaping subsequent patterns of reliance, acceptance, and delegation. This shifts the analytical focus from individual attitudes towards AI to the interactions through which these relationships develop and acquire broader social consequences. Repeated patterns of trust, fairness, and legitimacy can influence how roles, responsibilities, and influence are distributed within hybrid intelligence systems and how these arrangements become embedded in organisational structures. This relational perspective also provides a basis for examining how authority and leadership emerge in human–machine systems.

## 5.2 Authority, Leadership, and Organisational Integration

As AI systems become integrated into organisations, questions of trust and legitimacy intersect with questions of authority (Jarrahi, 2018; Glikson & Woolley, 2020). Trust concerns expectations about a system's competence and reliability, while reliance concerns whether and how people use its outputs. Authority concerns the decision rights and expectations of compliance attached to those outputs through organisational arrangements; legitimacy concerns whether these arrangements are regarded as appropriate. These dimensions need not coincide: employees may be required to follow a system's recommendations despite distrusting it, or trust its advice without accepting its right to determine their actions. AI use in recruitment, resource allocation, performance evaluation, customer service, and strategic planning therefore raises questions about how decision rights, discretion, and accountability are organised.

Contemporary AI systems increasingly perform functions associated with advisers, collaborators, coordinators, and, in some cases, managers. When AI becomes involved in tasks that have traditionally relied on human judgement, recommendations and decisions may be produced through processes that differ from human reasoning (Jarrahi, 2018). This difference matters from a Hybrid Intelligence perspective, which considers how distinct human and machine capabilities combine within shared decision processes (Dellermann et al., 2019). In organisational settings, machine-generated recommendations and decisions can enter existing relations of authority as humans rely on, contest, modify, or override them. Authority in these settings develops through how people interpret, act on, and incorporate machine-generated decisions into organisational practice.

Leadership evaluations are themselves shaped by anthropomorphism and social categorisation. People readily infer personality, warmth, competence, and authority from names, voices, avatars, communication styles, and other social cues embedded within AI systems. As a result, stereotypes commonly applied to human leaders may also transfer to artificial agents. Recent experimental evidence demonstrates that gender stereotypes influence perceptions of AI managers, with otherwise identical systems receiving different evaluations depending on whether they are presented as male or female (Cui & Yasseri, 2026). These studies demonstrate that perceptions of authority can depend on how an artificial agent is socially interpreted, even when its underlying algorithmic capabilities remain unchanged.

The distribution of authority depends on how machine outputs enter organisational decision procedures. Consider an illustrative case in which an AI recommends how work should be allocated. In one arrangement, staff can question its recommendations and a human manager decides whether to adopt them. In another, the same recommendations become default assignments that employees must justify overriding. The system's technical capabilities need not change for its organisational influence to increase. What changes is the arrangement governing compliance, discretion, and contestation. This example shows why technical competence, willingness to rely on an output, and authority within an organisation require separate analysis.

Research on shared mental models highlights the importance of expectations about capabilities, limitations, and responsibilities in human–AI teams (Andrews et al., 2023). A behavioural sociological analysis additionally asks who authorises a system's use, whose interests its decisions serve, and who can challenge them. Machine-generated outputs can acquire authority through organisational rules without the system holding authority in the same sense as a human role-holder. Responsibility remains with the human and organisational actors who establish, operate, and oversee these arrangements. For HI, role design consequently involves both coordinating capabilities and specifying the decision rights attached to their use.

From a behavioural sociological perspective, organisations are important settings in which human–machine relationships become routinised. They show how authority, leadership, and responsibility are negotiated when machine outputs enter established work practices. Hybrid Intelligence in these settings depends on how human and artificial capabilities are coordinated through roles, routines, and decision rights. Organisational practices may also diffuse into other domains of social life, but the extent and consequences of that diffusion require empirical investigation.

## 5.3 Norm Formation and Institutional Adaptation

As intelligent systems become embedded in everyday social life, their influence can extend beyond individual behaviour and organisational practice to shared norms and institutions.

Repeated interactions may contribute to shared expectations, routines, and formal rules when their effects are reinforced across encounters and incorporated into organisational practices. Human–machine interaction may contribute to this process, but behavioural carryover alone does not establish institutionalisation. Whether interaction patterns diffuse across communities and become accepted ways of communicating, coordinating, making decisions, and allocating responsibility remains an empirical question.

The literature suggests that AI contributes to norm formation in several ways. Intelligent systems may reinforce existing norms by consistently rewarding particular behaviours, discourage others through automated monitoring and moderation, or introduce entirely new expectations regarding communication, coordination, and decision-making. Recommendation systems shape the information people encounter, content moderation algorithms define acceptable online behaviour, and generative AI increasingly mediates writing, programming, education, and creative work. Through these processes, AI can contribute to how social norms are produced, stabilised, and disseminated (Taddeo & Floridi, 2018).

Norm formation, however, is not a one-way process in which technology reshapes society. Human users continually adapt their behaviour in response to intelligent systems, while AI itself is modified through user feedback, organisational practices, regulatory interventions, and retraining on new data. This reciprocal process reflects the sociotechnical perspective introduced earlier in the chapter, which holds that technologies and societies continually co-evolve rather than determining one another independently (Bijker et al., 1987; Latour, 2005).

The integration of AI also generates tensions between established and emerging norms. Questions of delegation, accountability, transparency, privacy, and responsibility frequently require organisations to develop new procedures governing how AI-generated recommendations should be interpreted, contested, and incorporated into decision-making. Changes to coordination can also challenge existing expectations of reciprocity, as illustrated by the semiautomatic driving experiments of Shirado et al. (2023).

Artificial agents may also become embedded within institutional arrangements that regulate collective behaviour. Experiments show that rewards and punishments can reduce biases against cooperation partners assumed to be machines (Makovi et al., 2025), while computational modelling identifies agent policies that may support cooperation under specified conditions (Hintze & Adami, 2026). These findings point to mechanisms for designing hybrid arrangements; they do not by themselves demonstrate the emergence of enduring institutions. Institutional reliance on algorithmic systems may also reproduce or amplify inequalities when biased outputs become embedded within routine organisational practice (Green & Chen, 2019; Selbst et al., 2019).

From a behavioural sociological perspective, these developments demonstrate that the societal consequences of AI emerge from the institutional contexts in which human–machine interactions take place. These systems become socially significant as they reshape the norms governing cooperation, authority, responsibility, and collective decision-making. Understanding these processes requires analysing institutions as evolving sociotechnical systems in which humans and intelligent machines continually adapt to one another. This institutional perspective completes the progression developed throughout the chapter, from individual perception and interaction to the emergence of collective behaviour, organisational practice, and broader social change.

## 5.4 Designing and Governing Human–Machine Systems

The literature reviewed throughout this chapter demonstrates that designing effective human-machine systems requires considerably more than improving the technical capabilities of AI models. Behavioural outcomes emerge through interactions among algorithms, users, organisations, and institutions, making successful human-machine systems fundamentally sociotechnical. Consequently, technical performance alone cannot guarantee trust, fairness, cooperation, legitimacy, or long-term social acceptance.

One important implication concerns transparency and explainability. Explanations can improve users' understanding of AI recommendations, support appropriate trust calibration, and strengthen accountability (Lee & See, 2004). However, transparency alone is rarely sufficient. People interpret explanations through existing beliefs, professional routines, organisational practices, and social expectations. Explainability develops through the interaction between the information an AI system provides and how users interpret and use it. Effective explanations need to account for how information is communicated, who receives it, and how it is incorporated into collective decision-making (Glikson & Woolley, 2020).

A second implication concerns fairness, responsibility, and governance. Research increasingly demonstrates that fairness cannot be achieved solely through technical interventions such as dataset balancing, bias mitigation, or algorithmic optimisation. As discussed earlier in this chapter, bias emerges throughout the sociotechnical system, from historical data and model design to interface design, organisational routines, institutional practices, and patterns of human interpretation (Selbst et al., 2019; Suresh & Guttag, 2021). Questions of responsibility consequently extend across the actors involved in designing, deploying, regulating, and using AI systems. Governance needs to address this distribution of responsibility through technical auditing, institutional oversight, mechanisms for contesting decisions, continuous monitoring of social outcomes, and clear accountability for decisions made jointly by humans and intelligent systems (Green & Chen, 2019; Raji et al., 2020).

Behavioural sociology treats humans and intelligent systems as interacting components of shared social systems. These systems develop through repeated interactions among human behaviour, machine behaviour, and the social and institutional settings in which they are embedded. Trust, cooperation, authority, legitimacy, and institutional change can emerge from these interactions and evolve as patterns of behaviour and social arrangements change over time. This approach provides a basis for examining how organisations and societies combine human and machine capabilities, and how their interaction affects collective outcomes such as effectiveness, fairness, resilience, and legitimacy.

The evidence reviewed throughout this chapter (summarised in Table 2) suggests that the future of human-machine systems will depend as much on behavioural and institutional design as on advances in artificial intelligence itself. Designing successful hybrid systems requires integrating technical innovation with insights from sociology, psychology, organisational science, and behavioural research. More fundamentally, it requires recognising that AI becomes socially consequential not because of what algorithms can do in isolation, but because of how humans and intelligent systems interact, cooperate, coordinate, and adapt over time. This behavioural and sociological perspective provides the central contribution of the emerging sociology of humans and machines.

Table 2. Behavioural domains in human-machine systems across levels of social organisation and their implications for design and governance.

| Behavioural domain | Individual level | Organisational and system level | Design and governance implication |
|---|---|---|---|
| Anthropomorphism | Attribution of human qualities and expectations | Trust calibration and role expectations | Communicate capabilities and limitations clearly |
| Social categorisation | Assigned identity and status expectations | Unequal evaluations and distributions of influence | Evaluate identity cues and organisational role design |
| Bias | Stereotyping and unequal judgements | Discriminatory organisational outcomes | Auditing, oversight, and accountability |
| Cooperation | Appropriate reliance and reciprocal behaviour | Human–AI teamwork and coordination | Role design and incentive structures |
| Authority | Deference to or contestation of machine outputs | Organisational delegation and decision rights | Accountability, contestability, and human oversight |
| Norm formation | Behavioural adaptation and shared expectations | Organisational routines and institutional reinforcement | Monitor persistence, diffusion, and social outcomes |

# 6. Conclusion

The preceding sections have traced human–machine interaction from individual perceptions and behavioural responses to collective, organisational, and institutional processes. Viewed through the lens of Hybrid Intelligence, these processes raise questions about how human and machine behaviour become interdependent and how this interdependence acquires broader social consequences. This brings the relationship between behavioural sociology and Hybrid Intelligence into focus. What can behavioural sociology contribute to the study of Hybrid Intelligence, and what does Hybrid Intelligence mean for behavioural sociology? This section addresses these questions and considers the challenges they raise.

## 6.1 Behavioural Sociology as a Perspective on Hybrid Intelligence

Behavioural sociology contributes to Hybrid Intelligence by examining the social processes through which human and machine capabilities become interdependent. Combining these capabilities does not necessarily make them complementary, as their behavioural consequences also depend on human responses and the social settings in which interaction takes place. This interdependence is shaped by how people interpret machine behaviour, how reliance and responsibilities are distributed, and how interaction is organised through roles, expectations, and norms.

Behavioural sociology also contributes to Hybrid Intelligence by tracing how human–machine interactions develop into broader collective and institutional processes. Repeated interactions can reshape coordination, distributions of influence and responsibility, organisational routines, and shared norms. These arrangements can then shape subsequent human–machine interactions. This cross-level perspective highlights how interdependence develops across groups, organisations, and institutions, and the collective consequences that follow.

## 6.2 What Hybrid Intelligence Means for Behavioural Sociology

Hybrid Intelligence changes the analytical object of behavioural sociology by making social outcomes depend on interactions among actors with different forms of capability and agency. The evidence and theoretical perspectives reviewed here suggest several consequences for established mechanisms. Social categorisation can assign human status expectations to designed identities; reciprocity can combine cooperative machine behaviour with unequal human obligations to reciprocate; and authority can attach to outputs through organisational procedures rather than through a machine's own understanding of its role. Machine participation may also change subsequent relationships among humans, as illustrated by behavioural effects that persist after automated assistance is deactivated. Explaining these outcomes requires tracing both human interpretations and the machine behaviours and institutional arrangements to which they respond.

These mechanisms can remain recognisable while changing in their operation, meaning, or distribution of obligations and influence. Observable cooperation does not establish shared intentions, and human deference to an algorithm does not by itself establish that the algorithm is a legitimate authority. A behavioural sociology of HI therefore needs to distinguish what a system does, what people believe it is doing, and how the surrounding organisation enables or constrains their responses. This provides a basis for examining when the combination of human and machine capabilities produces complementarity, and when its social organisation undermines that potential.

## 6.3 Outstanding Challenges

Beyond these conceptual questions, important empirical and methodological challenges remain. Much of the existing literature still focuses on dyadic interactions between a single human and a single AI system, whereas contemporary human-machine environments increasingly involve multiple humans interacting with multiple intelligent agents simultaneously. Understanding these larger and more complex systems requires greater attention to networks, organisations, institutions, and emergent collective behaviour.

Methodologically, controlled experiments remain central to behavioural research. They can identify causal effects under specified conditions, but their generalisability across tasks, populations, and institutional settings requires further evaluation. Future research should increasingly combine controlled experiments with longitudinal field studies, digital trace data, computational simulations, and natural experiments to examine behavioural dynamics as they unfold in real-world human-machine systems.

Substantively, many important questions remain open. We still know relatively little about how trust evolves over long periods of interaction, how stable human-machine norms emerge, how authority and responsibility should be distributed within hybrid organisations, or how AI-mediated interaction will influence inequality, social cohesion, democratic governance, and institutional legitimacy. Addressing these questions will require moving beyond studies of isolated human–AI interactions towards longitudinal, multilevel analyses of hybrid social systems in which intelligent machines become persistent participants in everyday social life.

## 6.4 Looking Forward

The behavioural sociology of humans and machines remains an emerging field, but its importance will continue to grow as intelligent systems become more deeply embedded within society. Future research is likely to move beyond asking whether AI performs better than humans towards understanding how humans and intelligent systems can jointly create more effective, equitable, resilient, and trustworthy forms of collective organisation. Achieving this goal will require continued collaboration across sociology, psychology, computer science, behavioural science, organisational research, and collective intelligence.

The broader implication of this chapter is that advances in artificial intelligence have social consequences that extend beyond technological development. They reshape how people cooperate, coordinate, communicate, organise institutions, and govern collective life. Viewed through the lens of Hybrid Intelligence, explaining these transformations requires attention to how human and machine behaviour becomes interdependent and how these interactions develop into broader collective, organisational, and institutional processes. A behavioural sociology of humans and machines offers a perspective for understanding these processes and their social consequences.

# Suggested Readings

**The Psychology of Human-Computer Interaction**
Card, S. K., Moran, T. P., & Newell, A. (1983). *The Psychology of Human-Computer Interaction*. Lawrence Erlbaum Associates.
A classic work that established the cognitive foundations of human-computer interaction and remains essential for understanding how people interact with computational systems.

**Machine Behaviour**
Rahwan, I., Cebrian, M., Obradovich, N., Bongard, J., Bonnefon, J.-F., Breazeal, C., et al. (2019). *Machine behaviour*. *Nature, 568*(7753), 477–486.
A landmark perspective proposing that intelligent machines should be studied as behavioural entities, laying the foundations for systematic research on machine behaviour.

**A New Sociology of Humans and Machines**
Tsvetkova, M., Yasseri, T., Pescetelli, N., & Werner, T. (2024). *A new sociology of humans and machines*. *Nature Human Behaviour, 8*(10), 1864–1876.
Introduces the emerging sociological perspective that increasingly autonomous AI systems should be understood as participants in social systems rather than merely as technologies acting upon them.

**Social Cognition: From Brains to Culture**
Fiske, S. T., & Taylor, S. E. (2013). *Social Cognition: From Brains to Culture* (2nd ed.). Sage.
A comprehensive introduction to social cognition that provides the theoretical foundations for understanding anthropomorphism, social categorisation, stereotyping, trust, and interpersonal judgement in human-machine interaction.

**Automation, Algorithms, and Beyond: Why Work Design Matters More Than Ever in a Digital World**
Parker, S. K., & Grote, G. (2022). Automation, algorithms, and beyond: Why work design matters more than ever in a digital world. *Applied Psychology, 71*(4), 1171–1204.
Provides an organisational perspective on how automation and AI reshape work design, including autonomy, skill use, task allocation, and the social conditions under which new technologies are integrated into work.

**AI-Teaming: Redefining Collaboration in the Digital Era**
Schmutz, J. B., Outland, N., Kerstan, S., Georganta, E., & Ulfert, A. S. (2024). AI-teaming: Redefining collaboration in the digital era. *Current Opinion in Psychology, 58*, 101837.
Provides an overview of human–AI teaming, with particular attention to communication, coordination, trust, shared cognition, and the processes that shape collaboration and performance in human–AI teams.

**Exploring AI-in-the-Making: Sociomaterial Genealogies of AI Performativity**
Scott, S. V., & Orlikowski, W. J. (2025). Exploring AI-in-the-making: Sociomaterial genealogies of AI performativity. *Information and Organization, 35*(1), 100558.
Offers a sociomaterial perspective on AI that examines how AI technologies and social practices develop through their ongoing relations, providing a complementary theoretical approach to understanding AI as part of organisational and social processes.

# Acknowledgement

Taha Yasseri and Hao Cui acknowledge support from the Irish Research Council Laureate Award, grant number IRCLA/2022/3217, ANNETTE (Artificial Intelligence Enhanced Collective Intelligence). Taha Yasseri also acknowledges support from Workday, Inc.

# AI Use Statement

During the preparation and revision of this chapter, ChatGPT (OpenAI) was used to assist with language editing, structural revisions, drafting and revising selected passages, and checks of conceptual consistency and bibliographic details. The authors evaluated and revised the AI-assisted material and remain fully responsible for the accuracy, originality, interpretation, and conclusions of the chapter.